\documentclass[a4paper]{article}

\usepackage{18lomcon}        
\usepackage{cite}             
\usepackage{epsfig}           
\usepackage{epstopdf}

\begin{document}


\title{THE T2HKK EXPERIMENT AND NON-STANDARD INTERACTION}

\author{Monojit Ghosh \email{mghosh@phys.se.tmu.ac.jp}
        and
        Osamu Yasuda \email{Yasuda@phys.se.tmu.ac.jp}
}

\affiliation{Department of Physics, Tokyo Metropolitan University, Hachioji, Tokyo 192-0397, Japan}


\date{}
\maketitle


\begin{abstract}
   In this work we study the the sensitivity of
   the T2HKK experiment to probe non-standard interaction in neutrino propagation. As this experiment will be statistically dominated due to its large detector volume and high beam-power, 
   it is expected that the sensitivity will be affected by systematics.
   This motivates us to study the effect of systematics in probing the non-standard interaction. 
   We also compare our results with the other future proposed experiments i.e., T2HK, HK and DUNE.
\end{abstract}

\section{Introduction}

In the standard three flavour framework, the phenomenon of neutrino oscillation where neutrinos evolve from one flavour from another can be parameterized by six oscillation parameters: 
three mixing angles: $\theta_{12}$, $\theta_{13}$ and $\theta_{23}$, two mass squared differences: $\Delta_{21}$ ($m_2^2 - m_1^2$) and $\Delta_{31}$ ($m_3^2 - m_1^2$) and one phase $\delta_{CP}$.
Among them at present the unknowns are: (i) sign of $\Delta_{31}$ i.e., $\Delta_{31} > 0$ (normal hierarchy or NH) or $\Delta_{31} < 0$ (inverted hierarchy or IH), 
(ii) octant of $\theta_{23}$ i.e., $\theta_{23} > 45^\circ$ (higher octant or HO) or $\theta_{23} < 45^\circ$ (lower octant or LO) and (iii) $\delta_{CP}$. 
T2HKK is one of the proposed experiment to determine these unknowns at a higher confidence level. 
In T2HKK \cite{Abe:2016ero}, there will be one water cerenkov detector of volume 187 kt in Kamioka 
and another 187 kt similar detector in Korea. Depending on the location in Korea there are different off-axis (OA) flux possibilities. In this similar context, other future oscillation experiments 
are T2HK (in which both the detector will be at Kamioka), HK (the atmospheric counterpart of T2HK) and 
DUNE \cite{Acciarri:2015uup} (the future project in Fermilab). Apart from determining the unknown oscillation parameters,
these experiments also give us opportunity to study new physics scenarios: for example non-standard interaction (NSI) which we will discuss in the next section.

\begin{figure*}[t]
\centering
\vspace{-1.0 in}
\includegraphics[scale=0.7]{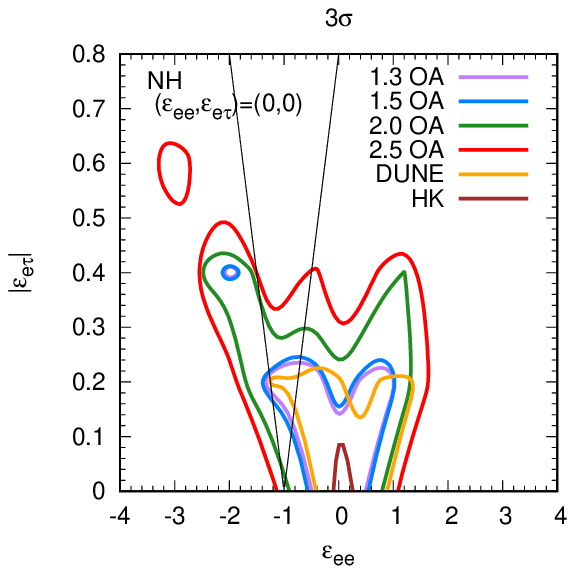}
\hspace{-50pt}
\includegraphics[scale=0.7]{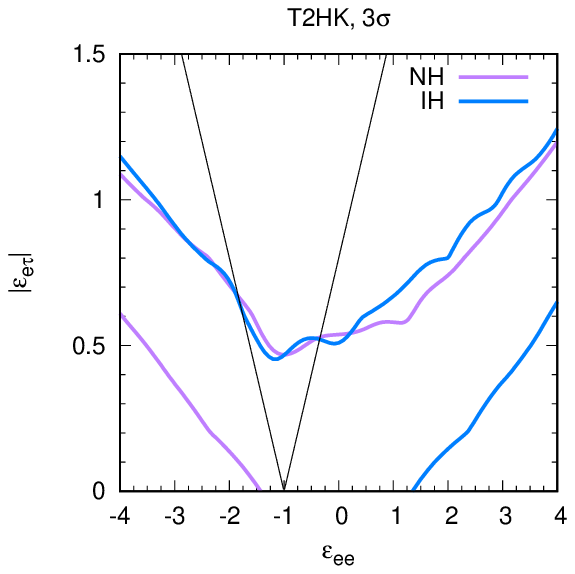} \\
\includegraphics[scale=0.7]{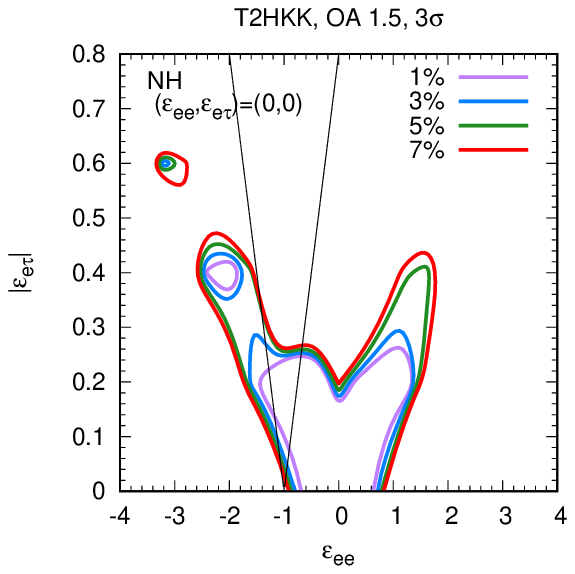}
\hspace{-50pt}
\includegraphics[scale=0.7]{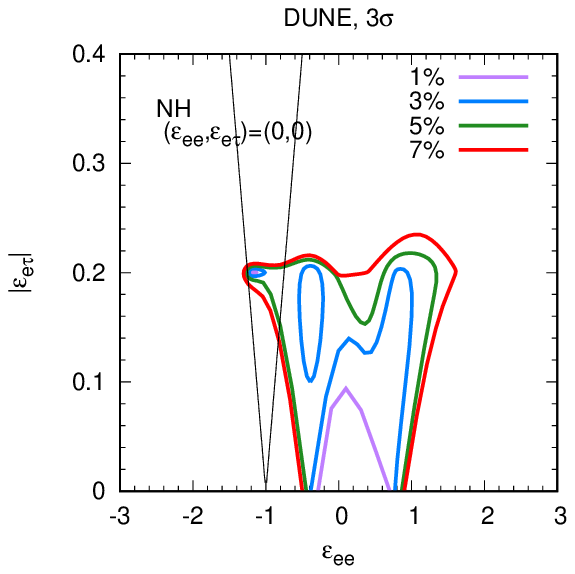}
\vspace{9 mm}
\caption{
The excluded region in the
($\epsilon_{ee}$, $|\epsilon_{e\tau}|$) plane. The thin solid diagonal straight line
stands for the bound $|\tan\beta|\equiv
|\epsilon_{e\tau}/(1+\epsilon_{ee})| < 0.8$.}
\label{fig1}
\end{figure*}


\section{Non-standard interaction}

Existence of NSI implies, the initial and final flavour of the neutrinos during the neutral current (NC) interaction with matter can be different \cite{Ohlsson:2012kf}. 
In this case, the matter term is modified by
\begin{eqnarray}
{\cal A} \equiv
\sqrt{2} G_F N_e \left(
\begin{array}{ccc}
1+ \epsilon_{ee} & \epsilon_{e\mu} & \epsilon_{e\tau}\\
\epsilon_{\mu e} & \epsilon_{\mu\mu} & \epsilon_{\mu\tau}\\
\epsilon_{\tau e} & \epsilon_{\tau\mu} & \epsilon_{\tau\tau}
\end{array}
\right),
\label{matter-np}
\end{eqnarray}
where $\epsilon_{\alpha\beta}$ are the NSI parameters.
The present $90\%$ bounds of the NSI parameters are given by\,\cite{Biggio:2009nt} 
\begin{eqnarray}
\hspace{-25pt}
&{\ }&
\left(
\begin{array}{lll}
|\epsilon_{ee}| < 4 \times 10^0 & |\epsilon_{e\mu}| < 3\times 10^{-1}
& |\epsilon_{e\tau}| < 3 \times 10^0\\
&  |\epsilon_{\mu\mu}| < 7\times 10^{-2}
& |\epsilon_{\mu\tau}| < 3\times 10^{-1}\\
& & |\epsilon_{\tau\tau}| < 2\times 10^1
\end{array}
\right).
\label{epsilon-m}
\end{eqnarray}
Thus we understand that the bounds on $\epsilon_{\alpha \mu}$ where $\alpha = e$, $\mu$, $\tau$ are stronger than the $\epsilon_{ee}$, $\epsilon_{e\tau}$ and $\epsilon_{\tau\tau}$. 
One additional bound comes from the high-energy atmospheric data which relates the parameters $\epsilon_{\tau\tau}$ and $\epsilon_{e \tau}$ as \cite{Friedland:2005vy}
\begin{eqnarray}
\epsilon_{\tau\tau} \simeq \frac{|\epsilon_{e\tau}|^2}{1+\epsilon_{ee}}\,.
\label{eq:ansatz_a}
\end{eqnarray}
Keeping these facts in mind, we perform our analysis with $\epsilon_{\alpha \mu} = 0 $.
Thus the free parameters are $\epsilon_{ee}$, $|\epsilon_{e \tau}|$ and 
arg($\epsilon_{e\tau}) = \phi_{31}$. In our analysis we have kept the true values of ($\delta_{CP}$, $\theta_{23}$) fixed at ($-90^\circ$, $45^\circ$).

\section{Results}
We have done our simulation using GLoBES \cite{Huber:2004ka} and MonteCUBES \cite{Blennow:2009pk}. 
In Fig. \ref{fig1}, we have given our results in the test $\epsilon_{ee}$ - test $|\epsilon_{e\tau}|$ plane. The true value of $\phi_{31}$ is zero and marginalized in test. 
For systematic errors, we have considered normalization error which affects the scaling of the events and tilt error which affects the energy dependence of the events.  
The tilt error is taken as 10\% and for the upper panels the normalization errors are taken from Refs. \cite{Abe:2016ero,Acciarri:2015uup}. For the bottom panels, we have given our results for
four different values of normalization errors.
From the left panel we see that among the different off-axis configurations of T2HKK, the sensitivity of
OA $1.3^\circ$ is best and comparable to DUNE. But the best sensitivity comes from HK. From the right panel we see that compared to T2HKK, the sensitivity of T2HK is quite weak.
From the bottom panels we see that the results depend on the systematic uncertainties significantly. 

\section{Summary}

In this work we have studied the sensitivity to NSI for T2HKK and compared our results with T2HK, HK and DUNE. We have also studied the effect of systematic uncertainties.
For more details, we refer to \cite{Fukasawa:2016lew,Ghosh:2017ged} on which this work is based upon.

\section*{Acknowledgments}

This research was partly supported by a Grant-in-Aid for Scientific
Research of the Ministry of Education, Science and Culture, under
Grants No. 25105009, No. 15K05058, No. 25105001 and No. 15K21734.


\end{document}